\documentclass[12pt]{article}

\usepackage{amsmath}
\usepackage{amsfonts}
\newcommand\bm[1]{\mbox{\boldmath$#1$}}

\def\lie{\mathcal{L}}

\def\ee{\varepsilon}

\def\gback{g^{(0)}}
\def\domega{\Omega_{AB}d\theta^A d\theta^B}
\def\gfpert{{g^{(1)}}}
\def\gTpert{g_{pert}}
\def\sup{\Sigma}
\def\supo{\Sigma_0}
\def\domegas{\Omega_{AB}|_{\supo} d\vartheta^A d\vartheta^B}
\def\embedo{Z^{(0)}}
\def\fff{q}
\def\sff{K}
\def\nback{n^{(0)}}

\def\aback{a^{(0)}}
\def\vnback{\vec n^{(0)}}

\def\fnback{\bm n^{(0)}}
\def\fffback{{q^{(0)}}}
\def\ffffpert{{q^{(1)}}}
\def\sffback{{K^{(0)}}}
\def\sfffpert{{K^{(1)}}}
\def\K{\mathcal{K}}
\def\Journal#1#2#3#4#5#6{(#5) ``#6'' {#1} {\bf #2} #3#4}

\def\CQG{\em Class. Quantum Grav.}

\def\PRD{\em Phys. Rev. D }

\def\JMP{\em J. Math. Phys.}

\def\MPL{\em Mod. Phys. Lett.}

\def\PHLT{\em Phys. Lett.}

\def\eqq{\stackrel{\Sigma}{=}}
\begin{document}

\title{Linear perturbations of matched spacetimes: the gauge
problem and background symmetries}

\author{
Marc Mars\footnotemark[1], Filipe C. Mena$^\flat$ and
Ra\"ul Vera\footnotemark[3]\\
\footnotemark[1] Facultad de Ciencias, Universidad de Salamanca,\\
Plaza de la Merced s/n, 37008 Salamanca, Spain \\
$^\flat$Departamento de Matem\'atica, Universidade do Minho,\\
Campus de Gualtar, 4710 Braga, Portugal\\
$^\flat$Mathematical Institute, University of Oxford,\\
24-29 St Giles', Oxford OX1 3LB, UK \\
\footnotemark[3] Fisika Teorikoaren Saila, Euskal Herriko\\
Unibertsitatea, Apt. 644, Bilbao 48080, Basque Country, Spain }

\maketitle
\begin{abstract}
We present a critical review about the study of linear perturbations
of matched spacetimes including gauge problems. We analyse the
freedom introduced in the perturbed matching by the presence of
background symmetries and revisit the particular case of spherically 
symmetry in n-dimensions.
This analysis includes settings with boundary layers such as brane world models
and shell cosmologies. 
\end{abstract}  
\section{Introduction}
An important aspect in any geometric gravitational theory is the
analysis of how to match two spacetimes.  This is true in particular
for General Relativity and its perturbation theory.  Despite the
relevance and maturity of the matching theory one often finds papers
where the matching conditions are not properly used. Most of the
difficulties arise from the fact that the matching conditions are
imposed in specific coordinate systems in a manner which is not
completely coordinate independent. More specifically, matching two
spacetimes requires identifying the boundaries pointwise, and
sometimes this identification is done implicitly by fixing spacetime
coordinates, without paying enough attention to the fact that solving
the matching involves {\it finding} an identification of the boundary
and that this should not be fixed a priori.

In perturbation theory this problem also arises, and it gets
complicated by the fact that the fields to be matched (as the
perturbed metric) are gauge dependent. So, in addition to a priori
choices of identifications of the boundary, there is also the
problem that particular gauges are often used. It may be argued that
the matching theory must be gauge independent and therefore it can
be performed in any gauge. This is true, but only when due care is
taken to ensure that the choice of gauge does not restrict, a
priori, the perturbed identification of the boundaries.

A complete description of the linearized matching conditions
has been achieved only recently by Carter and Battye
\cite{carter_battye} and independently by Mukohyama
\cite{Mukohyama}. To second order, the matching conditions have been
recently found in \cite{MarcPertMatch}. Despite these papers, we
believe that some confusion still lingers in the field, in
particular with respect to the existing gauge invariant
formulations. The aim of this paper is to try to clarify these
issues. In order to do that, we will critically discuss some of the
approaches proposed in the literature trying to make clear which are
the implicit assumptions made and to what extent are they justified.

The first papers discussing the perturbed matching theory are, as
far as we know, the classic papers by Gerlach and Sengupta
\cite{gersen_oddmatch,gersen_evenmatch}.  However, as explained
below, their description of the perturbed matching theory contains
imprecisions, and we will therefore start discussing their approach
pointing out the difficulties they encounter. A first attempt to
justify the claims in \cite{gersen_oddmatch,gersen_evenmatch} is due
to Mart\'{\i}n-Garc\'{\i}a and Gundlach \cite{josemcarsten}, who propose a
different but nevertheless closely related set of linearized
matching conditions. Pointing out the implicit assumption made by
these authors will also help us to try to explain the subtleties
inherent to the perturbed matching theory.

In \cite{Mukohyama} the linearized matching conditions are described
for arbitrary backgrounds, perturbations and matching hypersurfaces,
and then applied to the case of two background spacetimes
with a high degree of symmetry,
namely those which admit a maximal group of isometries acting on
codimension two spacelike submanifolds (e.g. spherically symmetric
spacetimes). In order to simplify the matching conditions, Mukohyama
derives a set of matching conditions for so-called \emph{doubly gauge
invariants}. However, a gap arises in his final conclusions as the
presented set of conditions for doubly gauge invariant quantities
for the linearized matching of spacetimes are only shown  to be
\emph{necessary conditions}. Analysing sufficiency touches directly on the
issue we are trying to emphasize in this paper, so we devote one
section to clarify this point, where we show how
these conditions are, stricktly speaking, \emph{not sufficient}.
Since the matching conditions in
terms of doubly gauge invariants are widely used in the literature,
we consider important to close this gap. Moreover, the constructions of
gauge invariant quantities using spherical harmonic decompositons
leaves out the $l=0$ and $l=1$ sectors. We will discuss this issue and its
consequences.

The paper is organized as follows. We start by summarising the
perturbed matching conditions in Section \ref{sec:lin.match}, where
we also describe the gauge freedom involved. Then, the procedures
used in the classic papers \cite{gersen_oddmatch,gersen_evenmatch}
together with the justifications and further developments in
\cite{josemcarsten} are reviewed in Section
\ref{sec:stgaugeinvform}. Section \ref{sec:symm_freedom} focuses on the
consequences of the existence of symmetries in the background
configuration, which will have relevance in our final discussion. Section
\ref{sec:sphersymmbacks} has three subsections. The first one
is devoted to present briefly the
procedure and results discussed in \cite{Mukohyama}
particularised to the case of spherically symmetric backgrounds.
In the second subsection we analyse the sufficiency of the doubly gauge invariant
matching conditions in \cite{Mukohyama}. The last subsection is devoted to
the study of the freedom left in the perturbation of the
matching hypersurface once the metric perturbations have been fixed at
both sides. We finish with an appendix where explicit expressions for
the discontinuities of the perturbed second fundamental forms in the spherical
case are given. Some of these expressions are used in the main text.

\section{Linearized matching}
\label{sec:lin.match}

In this section we describe the gauge freedom involved in the
linearised spacetime matching and summarise the perturbed matching
conditions.

\subsection{Gauge freedom}
\label{sec:gauge-freedom}

The purpose of the matching theory is to 
construct a new spacetime out of two spacetimes $M^{\pm}$ with boundary
by {\it finding} a suitable diffeomorphism between the boundaries which allows
for their pointwise identification. In particular, the matched
spacetime cannot be thought to exist beforehand. 
Another aspect to bear in mind is that the matching conditions 
involve exclusively tensors on the identified boundary $\sup$ and hence
any coordinate system in $M^{\pm}$ is equally valid. This is well-known
but it is still source of confusion sometimes.

In perturbed matching theory, not only the metrics are perturbed but also
the matching hypersurfaces may be deformed. Furthermore,  as for the metric,
the ``deviation'' of the matching hypersurface is also a gauge dependent quantity. 
This can be best understood by viewing perturbations as $\varepsilon$-derivatives
(at $\varepsilon=0$) of a one-parameter family of
spacetimes $(M^{+}_{\ee}, g_{\ee}^{+})$ with boundary
$\sup^{+}_{\ee}$. It is convenient to embed $M^{+}_{\ee}$
within a larger manifold (without boundary) $V^{+}_{\ee}$ to clarify the discussion.
A priori, the manifolds $(M^{+}_\ee,g_\ee)$ are completely distinct so it
makes no direct sense to talk about $\ee$-derivatives. It is
necessary to identify first the different manifolds so that a single point $p$
refers to one point on each of the manifolds. 
Obviously, there are infinite ways
to identify the manifolds, all of them equally valid a priori. This freedom 
leads to the gauge dependence of the perturbed
metric (and of any other geometrically defined tensor).
The identification above may, or may not, map
the boundaries $\sup^{+}_{\ee}$ among themselves. A priori, a point in $\sup^{+}_{0}$
may be mapped, for $\ee \neq 0$, to a point on $\sup^{+}_{\ee}$, to a point interior 
to $M^{+}_{\ee}$ or to a point exterior to $M^{+}_{\ee}$ (within the extension $V^{+}_{\ee}$)
which is not part of the manifold. How can we then take derivatives with
respect to $\ee$ at those later points? Since
only derivatives at $\ee=0$ are needed, restricting to infinitesimal
values of $\ee$ entails no loss of generality. Then, if for some small
$\ee$, a point $q \in
\sup^{+}_{0}$ is mapped to the exterior of $M^{+}_{\ee}$,
it follows from differentiability with respect to $\ee$
that $q$ is mapped, for the reverse value $-\ee$, to a point 
interior to $M^{+}_{\ee}$. Thus, perturbations can be defined at the boundary
by taking one sided derivatives, i.e. to take limits $\ee \rightarrow 0$, with
a sign restriction on $\ee$ (c.f. \cite{MarcPertMatch} for an alternative discussion).

However, an important issue remains: How do we describe the
deformation of the boundary $\supo^+$? As a set of points each
boundary $\sup^{+}_{\ee}$ maps, with the above identification, into
a hypersurface of the background spacetime, which we call
$\hat{\sup}^{+}_{\ee}$. In general, this hypersurface will not
coincide with $\supo^+$ and may well touch it or cross it. This gives us an idea
of how the boundary is deformed, but only as a subset, not pointwise.
In order to know how the boundary actually moves within the background,
we need to prescribe a priori a pointwise 
identification of $\supo^+$ with  $\sup^{+}_{\ee}$.
This identification is completely different
and independent from the one
described above involving spacetime points, and involves only the
points on the boundaries. As before, there are infinitely many ways
to identify the boundaries, and this defines a second gauge freedom,
which involves objects intrinsically defined on the boundary. This
gauge freedom will be referred as \emph{hypersurface gauge}, as
opposed to the usual \emph{spacetime gauge} described above.

With both identifications chosen, the deformation of the boundary within the
background can already be described: 
Fix a point $q$ on the background boundary $\supo^+$. The identification
of the boundaries defines a point $q_{\ee}$
on $\sup^{+}_{\ee}$, for each $\ee$. The spacetime identification
takes this point $q_{\ee}$ and maps it into 
a point $\hat{q}_{\ee}$ of the background $M^{+}_{0}$
(perhaps after a sign restriction on $\ee$).  Obviously
$\hat{q}_{\ee}$ belongs to the perturbed hypersurface
$\hat{\sup}_\ee^+$. We have therefore not only a deformation of the
background hypersurface as a set of points, but also pointwise
information.  It only remains to take the tangent vector of the
curve $\hat{q}_{\ee}$ at $\ee =0$, i.e. $\vec{Z}^{+} = \frac{d
\hat{q}_{\ee}}{d \ee} |_{\ee=0}$ which encodes completely the
deformation of the matching hypersurface as seen from the
background spacetime. Two final remarks are in order: (i) 
$\vec{Z}^{+}$ is defined exclusively on $\supo^+$, no extension
thereof is defined or required and (ii)  $\vec{Z}^{+}$ depends on
both the spacetime and hypersurface gauges, since its defining curve is
constructed using both identifications.  However, decomposing 
$\vec{Z}^{+} = Q^{+} \vec{n}^{\,0}_{+} + \vec{T}^{+}$, where 
$\vec{n}^{\,0}_{+}$ is the unit normal of $\supo^+$ (assumed non-null anywhere)
and $\vec{T}^{+}$ is tangent to it, it turns
out that $Q^{+}$ depends on the spacetime gauge but not on the
hypersurface gauge. This is because changing the hypersurface gauge
reorganizes the points within each $\hat{\sup}^{+}_{\ee}$, but
cannot modify any of them as a set of points.

Tensors defined intrinsically on the boundaries $\sup^{+}_{\ee}$ are
completely unaffected by the spacetime identification, and are
therefore invariant under spacetime gauge transformations.  Recall that
the matching conditions involve only objects intrinsic
to the matching hypersurfaces. Since the perturbed matching conditions
are, formally, just their $\ee$-derivatives, it follows by construction
that the perturbed matching conditions must be 
gauge invariant under spacetime gauge transformations.
This may seem surprising at first sight since the
matching conditions must involve the perturbed metric, which is obviously gauge dependent.
However, the conditions turn out to be gauge independent because they also involve
the deformation vector $\vec{Z}^{+}$, which is spacetime gauge
dependent. This vector is therefore of fundamental importance 
and must be taken into account in any sensible
approach to the problem, as we shall see next.

\subsection{Matching conditions}
\label{sec:match-conditions}

Let $(M^{\pm}_0,g^{\pm}_0)$ be $n$-dimensional
spacetimes with non-null boundaries $\supo^{\pm}$.
Matching them requires an identification of the boundaries, i.e. a pair of embeddings
$\Phi_\pm:\; \supo \longrightarrow M^\pm_0$ with $\Phi_\pm(\supo) = \supo^{\pm}$,
where $\supo$ is an abstract copy of any of the boundaries. 
Let $\xi^i$ $(i,j,\ldots=1,\ldots,n-1)$
be a coordinate system on $\supo$. Tangent vectors to
$\supo^{\pm}$ are obtained by $ e^{\pm \alpha}_i =
\frac{\partial \Phi_\pm^\alpha}{\partial \xi^i}$
$(\alpha,\beta,\ldots=0,\ldots,n-1)$. There are also  unique (up to 
orientation) unit normal vectors $\nback_{\pm}{}^{\alpha}$ to the boundaries. 
We choose them so that if $\nback_{+}{}^{\alpha}$ points towards $M^{+}$ then
$\nback_{-}{}^{\alpha}$ points outside of $M^{-}$ or viceversa. The first and second fundamental
are simply $\fffback_{ij}{}^{\pm}\equiv e^{\pm \alpha}_i e^{\pm \beta}_j
\gback{}_{\alpha\beta}|_{{}_{\sup^\pm}}, \quad
\sffback_{ij}{}^{\pm}=-\nback_{\pm\alpha} e^{\pm
\beta}_i\nabla^\pm_\beta e^{\pm \alpha}_j$. The matching conditions (in the absence of shells) 
require the equality of the first and second fundamental forms on $\supo^{\pm}$, i.e.
\begin{equation}
  \fffback_{ij}{}^{+}=\fffback_{ij}{}^{-},~~~
  \sffback_{ij}{}^{+}=\sffback_{ij}{}^{-}.
\label{eq:backmc}
\end{equation}
Under a perturbation of the background metric
$\gTpert^\pm=\gback{}^\pm+\gfpert{}^\pm$ and of the matching hypersurfaces via
$\vec{Z}^{\pm} = Q^{\pm} \vec{\nback}_{\pm} + \vec{T}^{\pm}$, the matching conditions
will be perturbatively satisfied if and only if \cite{Mukohyama}
\begin{equation}
\ffffpert{}^+_{ij} = \ffffpert{}^-_{ij} , \hspace{2cm}
\sfffpert{}^+_{ij} = \sfffpert{}^-_{ij} ,
\label{eq:fpertmc}
\end{equation}
with
\begin{eqnarray}
\ffffpert_{ij}{}^\pm&=&
{\cal L}_{\vec T^\pm} \fffback_{ij}{}^{\pm}+
2 Q^\pm \sffback_{ij}{}^{\pm}+
e^{\pm\alpha}_i e^{\pm\beta}_j \gfpert_{\alpha\beta}{}^\pm,
\label{eq:pertfirstS}\\
\sfffpert_{ij}{}^\pm&=&
{\cal L}_{\vec T^\pm} \sffback_{ij}{}^{\pm}-\epsilon D_iD_j Q^\pm+
Q^\pm(-\nback_\pm{}^\mu \nback_\pm{}^\nu
R^{(0)\pm}_{\alpha\mu\beta\nu}e^{\pm\alpha}_i e^{\pm\beta}_j+
\sffback_{il}{}^{\pm}{\sffback}^{l}_j{}^{\pm})
\nonumber\\
&+&\frac{\epsilon}{2}\gfpert_{\alpha\beta}{}^\pm
\nback_\pm{}^\alpha \nback_\pm{}^\beta \sffback_{ij}{}^{\pm}-
\nback_\pm{}_\mu S^{(1)\pm\mu}_{\alpha\beta} e^{\pm\alpha}_i e^{\pm\beta}_j,
\label{eq:pertsecondS}
\end{eqnarray}
where $\epsilon= \nback_{\alpha} \nback{}^{\alpha}$,
$D$ is the covariant derivative of $(\sup,\fffback{}^{\pm})$ and $
{S^{(1)}}_{\beta\gamma}^{\pm\alpha}\equiv
\frac{1}{2} (
\nabla^{\pm}_\beta\, \gfpert{}^{\pm\alpha}_\gamma+$
$\nabla^{\pm}_\gamma\, \gfpert{}^{\pm\alpha}_\beta-
\nabla^{\pm \, \alpha}\, \gfpert{}^\pm_{\beta\gamma})$.\footnote{We
will abuse slightly the notation and refer 
to vectors on $\supo$ and their images on spacetime with the same symbol.
The meaning should be clear from the context.}

In these equations, $Q^{\pm}$ and $\vec{T}^{\pm}$ are a
priori unknown quantities and fulfilling the matching conditions
requires {\it showing} that two vectors $\vec Z^\pm$ exist such that
(\ref{eq:fpertmc}) are satisfied. The spacetime gauge freedom can be exploited to 
fix either or both vectors $\vec Z^\pm$ a priori, but this should be avoided (or
at least carefully analysed) if additional spacetime gauge choices are made, in order not
to restrict a priori the possible matchings. Regarding the hypersurface gauge, this can be used to
fix one of the vectors $\vec{T}^{+}$ or $\vec{T}^{-}$, but not both.

As already stressed the linearized matching conditions are by construction spacetime
gauge invariant (in fact each of the tensors $\ffffpert_{ij}{}^{\pm}$, 
$\sfffpert_{ij}{}^{\pm}$ is). Moreover, the set of conditions (\ref{eq:fpertmc}) 
are hypersurface
gauge invariant, provided the background is properly matched,
since \cite{Mukohyama} under such a gauge transformation given by 
the vector $\vec\zeta$ on $\supo$, $\ffffpert_{ij}$ transforms as
$\ffffpert_{ij}+\lie_{\vec\zeta \, }\fffback_{ij}$, and similarly for
$\sfffpert_{ij}$.

\section{On previous spacetime gauge invariant formalisms}
\label{sec:stgaugeinvform}
The first attempt to derive a general formalism for the matching
conditions in linearized gravity is, to our knowledge, due to
Gerlach and Sengupta \cite{gersen_oddmatch}. Their approach
is based on the description of the matching hypersurface $\sup$ as a
level set of a function $f$ defined on the spacetime. Assuming the level sets
$\{f = \mbox{const}\}$ to be timelike, a field of spacelike
unit normals is defined as
$n_{\mu} = ( g^{\alpha\beta} f_{,\alpha} f_{,\beta} )^{-1/2} f_{,\mu}$.
The unperturbed matching conditions correspond to the continuity everywhere
(in particular across $\sup$) of the tensors
\begin{equation}
  \label{eq:projector}
  \fff_{\alpha\beta}\equiv g_{\alpha\beta} - n_\alpha n_\beta, \hspace{2cm}
  \sff_{\alpha\beta}\equiv \fff_\alpha{}^\mu \fff_\beta{}^\nu \nabla_\mu n_\nu,
\end{equation}
which are the spacetime versions of the first and second fundamental forms introduced
above. Being $f$ defined everywhere, it makes sense to perturb it in order to describe the
variation of the matching hypersurface. Obviously, by perturbing $f$
one also perturbs $n_{\mu}$. The perturbed matching conditions
proposed in \cite{gersen_oddmatch} read
\begin{equation}
  \label{eq:cond_gersen}
  \fff_\mu{}^\alpha \fff_\nu{}^\beta\triangle(\fff_{\alpha\beta})^+=
  \fff_\mu{}^\alpha \fff_\nu{}^\beta\triangle(\fff_{\alpha\beta})^-,~~~~~
  \fff_\mu{}^\alpha \fff_\nu{}^\beta\triangle(\sff_{\alpha\beta})^+=
  \fff_\mu{}^\alpha \fff_\nu{}^\beta\triangle(\sff_{\alpha\beta})^-,
\end{equation}
where $\fff_{\beta}{}^{\alpha}$ is the projector onto $\Sigma$,
$\triangle$ stands for perturbation and $+$ and $-$ denote the
quantities as computed from either side of the matching hypersurface $\sup$.
These expressions involve the projections of the
perturbations of $q_{\alpha\beta}$ and $K_{\alpha\beta}$ onto $\Sigma$.
The need of considering only the projected components
is justified in \cite{gersen_oddmatch} since the matching
conditions need to be intrinsic to the matching hypersurfaces.
However, Gerlach and Sengupta themselves note that conditions
(\ref{eq:cond_gersen}) are not 
gauge\footnote{Throughout this section gauge will refer to
spacetime gauge. Hypersurface gauges will only appear briefly
towards the end of the section.} invariant.

Since the main interest in \cite{gersen_oddmatch,gersen_evenmatch} refers to
spherically symmetric backgrounds, this ``ambiguity'' is fixed in that case
by finding suitable gauge invariant combinations of
the linearized matching conditions, which turn out to give a correct set
of necessary perturbed matching conditions in spherical symmetry.
It should be stressed however, that the authors consider these gauge invariant subset
to be sufficient also, with no further justification.

We know from the discussion in Sect. \ref{sec:gauge-freedom}
above that (\ref{eq:cond_gersen})
cannot be correct as it leads to a set of gauge dependent conditions. 
Since, on the other hand the proposal (\ref{eq:cond_gersen}) may look plausible,
it is of interest to point out where, and in which sense, it fails to be correct.

The first source of problems comes from assuming that the matched spacetime
is given beforehand. Indeed, $q_{\alpha\beta}$ and $K_{\alpha\beta}$
are spacetime tensors and they can
only exist (and be continuous) once the matched spacetime is constructed. 
But this is precisely the purpose of the matching conditions, so the 
conditions become circular. Another aspect of the same problem
is that one can only talk about continuity once the pointwise
identification of the boundaries is chosen. But a level set of a function
defines only a set of points and not the way those points must be identified.
A third  instance of the same issue is that tensor components must be expressed in some
basis, e.g. a common coordinate system covering both sides of $\Sigma$. But again this
cannot be assumed a priori. It needs to be constructed.

Let us however mention that once
the pointwise identification of the boundaries is chosen, 
the use of spacetime tensors is allowed provided they are finally projected 
onto the hypersurface. In that sense, and when properly used, using spacetime indices
may simplify some calculations notably (see  
Carter and Battye, \cite{carter_battye} where this notation is used
to derive the perturbed matching conditions).

Besides this aspect (which already affects the background matching) the
perturbed equations (\ref{eq:cond_gersen}) suffer from one extra problem.
The perturbations $\triangle(\fff_{\alpha\beta})(p)$ and
$\triangle(\sff_{\alpha\beta})(p)$ at a point $p$ in the background
can be defined by taking $\varepsilon$-derivatives at fixed $p$ and $\ee=0$
of the corresponding tensors 
(defined by $g_{\alpha\beta}(\ee)$ and  $f_{\ee}$). For each value of $\ee$,
the matching conditions impose the continuity of 
$\fff_{\alpha\beta}(\ee)$ and $\sff_{\alpha\beta}(\ee)$
everywhere
(with the caveat already mentioned regarding the
identification of the boundaries). However, continuity of
$\triangle(\fff_{\alpha\beta})$ and
$\triangle(\sff_{\alpha\beta})$ at $p$ would only follow if derivatives
of continuous functions with respect to an external parameter were necessarily
continuous (in our case, the derivative with respect to $\ee$), which is not true in general. A trivial example is given
by the function $u(\ee,x) = | x+ \ee|$, whith $x \in \mathbb{R}$.
For each $\ee$ this function is continuous. However, the derivative
with respect to $\ee$ does not even exist at $x=0, \ee=0$.
This reflects the fact that subtracting continuous tensors at a fixed
spacetime point $p$ leads to objects that need not be continuous. This is in fact  the main
problem of (\ref{eq:cond_gersen}) as linearized matching conditions.

An immediate question arises: Why is the gauge invariant
subset of matching conditions found
in \cite{gersen_oddmatch,gersen_evenmatch}
for spherically symmetric backgrounds
correct? In order to understand this, let us rewrite (\ref{eq:cond_gersen})
using the formalism of section
\ref{sec:match-conditions}. 
First of all, since $\triangle (n_{\alpha} n_{\beta})$
will  contain, at least, one free $\nback_{\alpha}$, we have
\begin{equation}
\label{iden:fff}
 \fff_\mu{}^\alpha \fff_\nu{}^\beta\triangle(\fff_{\alpha\beta})^{\pm}=
  \fff_\mu{}^\alpha \fff_\nu{}^\beta \gfpert_{\alpha\beta}^{\pm}.
\end{equation}
Moreover, a simple calculation gives 
$\triangle (\nabla_{\alpha} n_{\beta} ) = \nabla_{\alpha}
( \triangle n_{\beta} ) - S^{(1)\mu}_{\alpha\beta} \nback_{\mu}$
and  $\triangle (\fff^{\alpha}_{\beta} )=
- \gfpert^{\alpha\mu} \nback_{\mu} \nback_{\beta} + \gback{}^{\alpha\mu}
\triangle (n_{\mu} ) \nback_{\beta} +
\nback{}^{\alpha} \triangle ( n_{\beta} )$. These, together with 
standard properties of the projector, lead to
\begin{equation}
\label{iden:sff}
  \fff_\mu{}^\alpha \fff_\nu{}^\beta\triangle(\sff_{\alpha\beta}) {}^{\pm}=
\left ( \aback_{\nu}\fff_\mu{}^\alpha \triangle (n_{\alpha}) +
\fff_\mu{}^\alpha \fff_\nu{}^\beta \nabla_{\alpha} ( \triangle n_{\beta} )
- \fff_\mu{}^\alpha \fff_\nu{}^\beta
S^{(1)\rho}_{\alpha\beta} \nback_{\rho} \right ) {}^{\pm},
\end{equation}
where $\aback_{\nu} \equiv \nback{}^{\alpha} \nabla_{\alpha} \nback_{\nu}$.
In general, these expressions do not agree with 
(\ref{eq:pertfirstS}) and (\ref{eq:pertsecondS}).
However, when the gauges are chosen so that $\vec{Z}^{\pm}=0$,
then $\triangle f \equiv 0$ on $\sup$ because the matching hypersurface is unperturbed
as seen from the background. Consequently 
$\partial_{\alpha} ( \triangle f) \propto \nback_{\alpha}$ on $\sup$, which
implies $\triangle (n_{\alpha}) \eqq h \nback_{\alpha}$ for some function
$h$. Imposing $\vec{n}(\ee)$ to be unit for all $\ee$ fixes
$h = \frac{\epsilon}{2}  \gfpert^{\alpha\beta} \nback_{\alpha} \nback_{\beta}$.
Inserting into (\ref{iden:sff}) the matching conditions
(\ref{eq:cond_gersen}) become
\begin{eqnarray}
\label{GSSurfaceGauge}
\left (
  \fff_\mu{}^\alpha \fff_\nu{}^\beta \gfpert_{\alpha\beta} \right )^{+} & = &
\left (
  \fff_\mu{}^\alpha \fff_\nu{}^\beta \gfpert_{\alpha\beta} \right )^{-},  \\
\left (
\frac{\epsilon}{2} \gfpert^{\alpha\beta} \nback_{\alpha} \nback_{\beta}
K_{\mu\nu} - \fff_\mu{}^\alpha \fff_\nu{}^\beta
S^{(1)\rho}_{\alpha\beta} \nback_{\rho} \right ) {}^{+} & = &
\left (
\frac{\epsilon}{2} \gfpert^{\alpha\beta} \nback_{\alpha} \nback_{\beta}
K_{\mu\nu} - \fff_\mu{}^\alpha \fff_\nu{}^\beta
S^{(1)\rho}_{\alpha\beta} \nback_{\rho} \right ) {}^{-} \nonumber,
\end{eqnarray}
which agree with (\ref{eq:fpertmc}) (with the exception that
(\ref{GSSurfaceGauge}) refers to spacetime tensors and (\ref{eq:fpertmc})
are defined on $\sup$).
Since Gerlach and Sengupta derive a subset of
gauge invariant matching conditions out of (\ref{eq:cond_gersen})
in the spherically
symmetric case and their conditions are correct in one gauge,
it follows that the invariant subset is correct in any gauge.
This is the reason why the results in \cite{gersen_oddmatch,gersen_evenmatch}
involving spherically symmetric backgrounds turn out to be fine.

Substantial progress in the linearized matching problem was made by
Mart\'in-Garc\'ia and Gundlach \cite{josemcarsten}.
These authors pointed out the lack of justification 
in \cite{gersen_oddmatch,gersen_evenmatch} for the choice of
(\ref{eq:cond_gersen}) as matching conditions. It was also argued
that for spacetimes with boundary 
it only makes sense to define perturbations
by using gauges where the perturbed matching hypersurface is mapped
onto the background matching hypersurface. Perturbations in this gauge,
called ``surface gauge'' (not to be confused with hypersurface gauge)
are denoted by $\bar\triangle$, and its defining property is
$\bar{\triangle}f = 0$. The idea
was to write down the matching conditions in this gauge and then
transform into any other gauge if necessary. As noticed by the
authors, the surface gauge is not unique since there are still three
degrees of freedom left, which correspond to the three directions tangent to
$\sup$.

A relevant observation made in \cite{josemcarsten} was that the
continuity of tensorial perturbations may depend on the index position
in the tensors. The authors argue that the tensors truly
intrinsic to the hypersurfaces are $\fff^{\alpha\beta}$,
$\sff^{\alpha\beta}$ (with indices upstairs) and propose
the following perturbed matching conditions
\begin{equation}
  \label{eq:cond_josemcarsten}
  \bar{\triangle}(\fff^{\alpha\beta})^+=\bar{\triangle}(\fff^{\alpha\beta})^-,
  ~~~~~
  \bar{\triangle}(\sff^{\alpha\beta})^+=\bar{\triangle}(\sff^{\alpha\beta})^-,
\end{equation}
which are demonstrated to become exactly (\ref{GSSurfaceGauge}). This shows the
equivalence of both proposals in the surface gauge, as explicitly stated in 
\cite{josemcarsten}. This justifies partially the validity of both approaches
in the surface gauge. However,
the justification is not complete because of the issue we discuss
next.

Indeed, conditions (\ref{eq:cond_josemcarsten}) still carry one
implicit assumption that needs to be clarified. As already stressed
the perturbed matching conditions have two inherent
and independent degrees of gauge freedom. The approach by
Mart\'{\i}n-Garc\'{\i}a and Gundlach involves only spacetime objects,
and therefore can only notice the spacetime gauge freedom.  This leads
to an incorrect statement in \cite{josemcarsten}, as it is not true that the linearized
matching conditions read (\ref{eq:cond_josemcarsten}) in \emph{any}
surface gauge. Conditions (\ref{eq:cond_josemcarsten}) 
will only be valid when the spacetime gauge maps pairs of
background points (identified, 
via the background matching) to pairs of points on the
perturbed boundaries $\Sigma^{\pm}_{\ee}$ which 
are also identified through the matching. Notice that not all
surface gauges have this property. In explicit terms, this means
that the vectors $\vec{Z}^{\pm}$ must (i) only have tangential
components (so that we are in surface gauge) and
(ii) have the same components when written in terms of an intrinsic basis of
$\supo$.
In less precise, but
more intuitive terms, condition (ii) states that $\vec{Z}^{+}$ and
$\vec{Z}^{-}$ are the same vector, i.e. that the gauges in both
regions are chosen such that the displacement of one fixed point of
the background hypersurface is identical in both regions (the
displaced point, of course, stays on the unperturbed hypersurface, due
to the choice of surface gauge). Observe finally that if $Q^{\pm}=0$
and $\vec{T}^{+} = \vec{T}^{-}$, then the linearized matching conditions
(\ref{eq:fpertmc}) truly reduce to conditions (\ref{GSSurfaceGauge}), once the latter
are projected on $\sup$. This shows the correctness of the approaches by 
Gerlach and Sengupta and Mart\'{\i}n-Garc\'{\i}a and Gundlach in special gauges.

\section{Freedom in matching due to symmetries}
\label{sec:symm_freedom}

We devote this section to the study of the consequences of the
existence of background symmetries on perturbed spacetime matchings.

The existence of symmetries in the background configuration
introduces two issues which are important to take into consideration:
the first corresponds to the freedom
introduced by the matching procedure, when preserving the symmetries,
at the background level \cite{mps}, c.f. \cite{MASEuni} for an application.
The second issue corresponds to the consequences
that the symmetries in the background configuration may have on the
perturbation of the matching.

It must be stressed here that the arbitrariness introduced
by the presence of symmetries in the background configuration
is completely independent from both the  hypersurface
and spacetime gauge freedoms.
However, that arbitrariness is gauge dependent and therefore
a gauge choice can be made to remove it. 
As we will show, an isometry in the background
implies that there is a direction along which the difference
$[\vec T]\equiv \vec T^+-\vec T^-$ 
cannot be determined by the perturbed matching equations.
But, as we have discussed at the
end of section \ref{sec:lin.match}, one could eventually choose
part of the spacetime gauges (if there is any freedom left)
to fix $[\vec T]$.
Note, finally, that a change of hypersurface gauge leaves $[\vec T]$ invariant.

\subsection{Isometries}

We shall now consider the presence of isometries in the background configuration.
So, let us assume that
one of the sides, say $(M^+_0,\gback{}^+)$,
admits an isometry generated by the Killing vector field $\vec\xi^+$
tangent to the boundary $\supo^+$. The commutation of the Lie derivative and the pull-back
implies \cite{mps}
\[
\lie_{\vec\xi^+}\fffback_{ij}{}^+=
e^{+\alpha}_i e^{+\beta}_j
\lie_{\vec\xi^+}\gback{}_{\alpha\beta}{}^+|_{\supo} =0,
\]
which means that 
$\vec\xi^+$ is a Killing vector of $(\supo,\fffback_{ij}{}^+)$.
This implies from expression (\ref{eq:pertfirstS})
that 
$\ffffpert{}_{ij}{}^+$ is invariant under the transformation
$\vec T^+\rightarrow \vec T^++\vec\xi^+|_{\supo}$.

As for $\sfffpert{}_{ij}{}^+$, from 
its expression (\ref{eq:pertsecondS}), it is again clear
that the previous transformations
of $\vec T^+$ will leave $\sfffpert{}_{ij}{}^+$
invariant provided
$\lie_{\vec\xi^+}\sffback{}_{ij}{}^+=0$.
But this is precisely the case since $\vec\xi^+$ is a Killing vector 
orthogonal to $\fnback_+$, which implies $
\lie_{\vec\xi^+}\fnback_+{}_\beta|_{\supo^{+}} =0$,
and hence
\[
\lie_{\vec\xi^+}\sffback{}_{ij}{}^+
=e^{+\alpha}_i e^{+\beta}_j
\lie_{\vec\xi^+} (\nabla\fnback_+)_{\alpha\beta}|_{\supo} =
e^{+\alpha}_i e^{+\beta}_j \nabla_\alpha
\lie_{\vec\xi^+}\fnback_+{}_\beta|_{\supo} =0.
\]
Of course, all this discussion also applies to the ($-$) side.

The combination of the invariance of $\ffffpert{}_{ij}{}^\pm$
and $\sfffpert{}_{ij}{}^\pm$ leads to the fact that the first order perturbed
matching conditions are invariant under a change
of the vectors $\vec T^\pm$ along the direction of any isometry
of the background configuration (preserved by the matching).
Then, as expected, when symmetries are present
the linearized matching conditions
cannot determine the difference $[\vec T]$
completely:
they leave undetermined the relative (between the two sides)
deformation of the hypersurface along the direction of the symmetry.
Note that, still, the \emph{shape} of the perturbed hypersurface
is completely determined, since that is driven by $Q^\pm$.

The overall picture is as follows:
at the background level we have the arbitrariness of the identification
of $\supo^+$ with $\supo^-$ \cite{mps}, which can be seen as a ``sliding''
between $\supo^+$ and $\supo^-$.
The perturbation adds  to this 
an arbitrary shift 
of the deformation of the matching hypersurface at 
each side along the orbits of the isometry group.
As an example, in the description of 
stationary and axisymmetric compact bodies 
discussed in \cite{MASEuni,mps},
the background sliding corresponds to an arbitrary
constant rotation of the interior with respect to the exterior.
Note that, in that case, this rotation is only relevant because the exterior is
taken to be asymptotically flat.
As a result, two identical interiors can, in principle,
give rise to two exteriors that differ 
by a constant rate rotation \cite{MASEuni}.
The shift of the surface deformation 
would, in principle, lead to an arbitrary constant 
rotation along the axial coordinate
of the surface deformation of the body.
Likewise, two identical perturbations in the interior of the body
may produce two different perturbations in the exterior,
which may differ by a relative constant rate rotation.
A choice of spacetime gauge could be used to relate
the deformations inside
and outside. However, this may
interfere with other gauge fixings that may have been made.

\section{$n$-dimensional spherically symmetric backgrounds}
\label{sec:sphersymmbacks}

In this section we shall revisit Mukohyama's theory for
linearized matching in the
special case of spherical symmetry. Similar results \cite{Mukohyama}
hold for 
backgrounds 
admitting isometry groups of dimension $(n-1)(n-2)/2$
acting on non-null codimension-two orbits of
arbitrary topology (strictly speaking the orbits need to be compact).

\subsection{The approach of Mukohyama}

Concentrating on one of the two spacetimes to be matched, either $+$ or $-$,
we consider a spherically symmetric background metric of the form
\begin{equation}
  \label{eq:ds2back}
  \gback_{\alpha\beta} dx^\alpha dx^\beta= \gamma_{ab} dx^a dx^b+
  r^2 \domega,
\end{equation}
where $\gamma_{ab}$ ($a,b,..=0,1$) is a Lorentzian two-dimensional metric
(depending only on $\{x^a\}$), $r>0$ is a function of $\{x^a\}$,
and $\domega$ is the $n-2$ dimensional unit sphere metric
with coordinates $\{\theta^A\}$ $(A,B,\ldots=2,3,\ldots,n-1)$.

A general spherically symmetric background hypersurface
can
be given in parametric form as
\begin{equation}
  \label{eq:sup0}
  \supo:= \{x^0=\embedo{}^0(\lambda),x^1=\embedo{}^1(\lambda),
  \theta^A=\vartheta^A\},
\end{equation}
where
$\{\xi^i\}=\{\lambda,\vartheta^A\}$ is a coordinate
system in $\supo$ adapted to the spherical symmetry.
The tangent vectors to $\supo$ read 
\begin{equation}
  \label{eq:tangent}
  \vec e_\lambda=
  \left. \dot \embedo{}^0\partial_{x^0}+\dot \embedo{}^1\partial_{x^1}
  \right|_{\supo},~~~
  \vec e_{\vartheta^A}=\left.\partial_{\theta^A} \right|_{\supo},
\end{equation}
where dot is derivative w.r.t. $\lambda$. With
$N^2\equiv -\epsilon e_\lambda{}^a e_\lambda{}^b\gamma_{ab}|_{\supo}$,
so that $\epsilon=1$ ($\epsilon=-1$) corresponds to a
timelike (spacelike) hypersurface,
the unit normal to $\Sigma_0$ reads
\begin{equation}
  \label{eq:normal}
  \fnback={\frac{\sqrt{-\det\gamma}}{N}}
  \left.\left(-\dot\embedo{}^1 d x^0+\dot\embedo{}^0 dx^1\right)\right|_{\supo},
\end{equation}
where the sign choice of $N$ corresponds to the choice of orientation of 
the normal.  The background induced metric and second 
fundamental form on $\supo$ read
\begin{eqnarray}
  \label{eq:fffback}
  \fffback_{ij}d\xi^i d\xi^j=-\epsilon N^2 d\lambda^2+r^2|_{\supo} \domegas, \\
  \label{eq:sffback}
  \sffback_{ij}d\xi^i d\xi^j=N^2\K d\lambda^2+
  r^2\bar\K|_{\supo}\domegas,
\end{eqnarray}
where
\[
\K\equiv N^{-2}e_\lambda{}^a e_\lambda{}^b\nabla_a \nback_b,~~~
\bar\K=\nback{}^a\partial_{x^a}\ln r.
\]
It follows that 
the background matching conditions (\ref{eq:backmc})
are 
\begin{equation}
  \label{eq:backmc2}
  N^2_+=N^2_-,~~~ r^2_+|_{\supo}=r^2_-|_{\supo},~~~
  \K_+=\K_-,~~~ \bar\K_+=\bar\K_-.
\end{equation}
Using (\ref{eq:pertfirstS}) and (\ref{eq:pertsecondS}) we could now compute
the first order perturbations 
$\ffffpert_{ij}$ and $\sfffpert_{ij}$
in terms of the above quantities and
$\vec Z$ (or equivalently $Q$ and $\vec T$), c.f. Eqs. (45) and (46) in \cite{Mukohyama}.
Let us recall 
(see subsection \ref{sec:match-conditions}) that 
while the individual tensors $\ffffpert_{ij}$ and $\sfffpert_{ij}$ are not
hypersurface gauge invariant, their respective differences from the $+$ and
$-$ sides (i.e.
the linearized matching conditions) are. Those tensors depend of the
hypersurface gauge 
through the tangent vectors $\vec{T}^{+}$ and $\vec{T}^{-}$, which 
under a gauge change transform simply by adding the gauge vector. 
It follows that only their difference
$[\vec{T}]$ 
can appear in the linearized matching conditions. Consequently
there are three  degrees of freedom that cannot be
fixed by the equations, but can be fixed by choosing
the hypersurface
gauge, for instance to set $\vec{T}^{+}$. Thus, the linearized matching
conditions can be looked at as equations for the difference $[\vec{T}]$ as
well as for $Q^{+}$ and $Q^{-}$, i.e. for five objects. If these
equations admit solutions, then the linearized matching is possible and it is impossible
otherwise. 

Mukohyama emphasizes the convenience to look for doubly gauge invariant
quantities to write down the linearized matching conditions, however
the matching conditions
are {\it already} gauge invariant (both for the spacetime
and hypersurfaces gauges). Looking for gauge invariant combinations on each side
amounts to writing equations where the difference vector $[\vec{T}]$ simply
drops. Indeed, in many cases, knowing the value of such vector in a specific  
matching is not interesting. 
In 
that sense, using doubly gauge invariant quantities is useful as it
lowers the number of equations to analyse. However, we want to stress that
this is not related to obtaining gauge invariant linearized matching equations. 
It is just related to not solving for superfluous information. In fact, a set
of equations where also $Q^{+}$ and $Q^{-}$ have disappeared would be even more
convenient from this point of view, provided one is not interested in knowing
how the hypersurfaces are deformed in the specific spacetime gauge being used.

Since the use of doubly gauge invariant matching conditions
is used extensively, let us recall its main ingredients in order to discuss
if they really are equivalent to the full set of linearized matching equations
and in which sense.

To that aim Mukohyama \cite{Mukohyama}, 
decomposes the perturbation tensors 
$\ffffpert_{ij}$ and $\sfffpert_{ij}$ in terms of
scalar $Y$, vector $V_A$  and tensor harmonics $T_{AB}$ on the sphere,
as\footnote{The ranges of $l$'s are not made explicit in \cite{Mukohyama}
in order to include also non-compact homogeneous spaces, where the index
$l$ is continuous. However, to discuss
sufficiency of the equations we need to be precise on the range of validity
of each equation.}
\begin{eqnarray}
  \label{eq:ffffpertharm}
  \ffffpert_{ij} d\xi^i d\xi^j&=&
  \sum_{l=0}^\infty (\sigma_{00} Y d\lambda^2+\sigma_{(Y)}T_{(Y)AB}
  d\vartheta^A d\vartheta^B)+
  \sum_{l=1}^\infty
  2(\sigma_{(T)0} V_{(T)A}+\sigma_{(L)0} V_{(L)A}) d\lambda d\vartheta^A \nonumber\\
  &&
  + \sum_{l=2}^\infty
    (\sigma_{(T)}T_{(T)AB}+
    \sigma_{(LT)}T_{(LT)AB}+\sigma_{(LL)}T_{(LL)AB}) d\vartheta^A d\vartheta^B,
\end{eqnarray}
\begin{eqnarray}
  \label{eq:sfffpertharm}
  \sfffpert_{ij}d\xi^i d\xi^j&=&
  \sum_{l=0}^\infty (\kappa_{00} Y d\lambda^2+\kappa_{(Y)}T_{(Y)AB}
    d\vartheta^A d\vartheta^B)+
  \sum_{l=1}^\infty
  2(\kappa_{(T)0} V_{(T)A}+\kappa_{(L)0} V_{(L)A}) d\lambda d\vartheta^A \nonumber\\
  &&
  + \sum_{l=2}^\infty
    (\kappa_{(T)}T_{(T)AB}+
    \kappa_{(LT)}T_{(LT)AB}+\kappa_{(LL)}T_{(LL)AB}) d\vartheta^A d\vartheta^B,
\end{eqnarray}
where all the scalar coefficients depend only on $\lambda$.
Each coefficient in the decomposition has indices $l$ and $m$ which have been
dropped for notational simplicity. Notice that each 
coefficient $\sigma$ and $\kappa$ is defined in the range 
of $l$'s appearing in the corresponding summatory.
By construction, each of the 
$\sigma$ and $\kappa$ 
are spacetime-gauge invariant (but not hypersurface-gauge invariant).
For $l\geq 2$ they can even be written down \cite{Mukohyama}
explicity in terms of spacetime-gauge invariant quantities. 
In a similar way,  the doubly gauge-invariant quantities 
presented in \cite{Mukohyama}, are only defined for $l\geq 2$
(except $k_{(T)0}$, which is also defined for $l=1$), and read
\begin{eqnarray}
  l\geq2:&&f_{00} \equiv \sigma_{00}-2 N \partial_\lambda\left(N^{-1}\chi\right),
  \nonumber\\
  l\geq2:&&f \equiv \sigma_{(Y)} +
  \epsilon N^{-2}\chi \partial_\lambda\left( r^2|_{\supo}\right)+
  \frac{2}{n-2}k_l^2\sigma_{(LL)},
  \nonumber\\
  l\geq2:&&f_0 \equiv \sigma_{(T)0}-r^2|_{\supo}\partial_\lambda\left(
  r^{-2}|_{\supo}\sigma_{(LT)}\right),
  \nonumber\\
  l\geq2:&&f_{(T)}\equiv \sigma_{(T)},
  \nonumber\\l\geq2:&&k_{00}\equiv \kappa_{00}+\epsilon\K\sigma_{00}+
  \epsilon\chi\partial_\lambda\K,
  \nonumber\\
  l\geq1:&&k_{(T)0} \equiv \kappa_{(T)0}- \bar\K\sigma_{(T)0},\label{eq:doublies}\\
  l\geq2:&&k_{(L)0} \equiv \kappa_{(L)0}
  +\frac{1}{2}(\epsilon\K-\bar\K)\sigma_{(L)0}
  +\frac{1}{2}(\epsilon\K+\bar\K)
  \left[
    \chi - r^2|_{\supo} \partial_\lambda(r^{-2}|_{\supo}\sigma_{(LL)})
  \right],
  \nonumber\\
  l\geq2:&&k_{(LT)} \equiv \kappa_{(LT)}-\bar\K\sigma_{(LT)},\nonumber\\
  l\geq2:&&k_{(LL)} \equiv \kappa_{(LL)}-\bar\K\sigma_{(LL)},\nonumber\\
  l\geq2:&&k_{(Y)} \equiv \kappa_{(Y)}-\bar\K\sigma_{(Y)}
  +\epsilon N^{-2} r^2|_{\supo}\chi\partial_\lambda\bar\K,\nonumber\\
  l\geq2:&&k_{(T)}\equiv \kappa_{(T)}-\bar\K \sigma_{(T)},\nonumber
\end{eqnarray}
where $k_l^2=l(l+n-3)$ and
\[
l\geq2:\hspace{1cm}\chi\equiv \sigma_{(L)0}-r^2|_{\supo}
\partial_\lambda(r^{-2}|_{\supo}\sigma_{(LL)}).
\]

The orthogonality properties of the scalar, vector and tensor
harmonics imply that the equalities
of the coefficients $\sigma$ and $\kappa$ for each $l$ and $m$
is equivalent to
the equality of the perturbation tensors (\ref{eq:ffffpertharm})
and (\ref{eq:sfffpertharm}) at both sides of $\supo$.
Thus, recalling the notation $[f]\equiv f^+|_{\supo}-f^-|_{\supo}$,
the equations
\begin{eqnarray}
  \label{eq:sigmes}
  &&\begin{array}{lll}
  l\geq 0:&&[\sigma_{00}]=[\sigma_{(Y)}]=0\\
  l\geq 1:&&[\sigma_{(L)0}]=[\sigma_{(T)0}]=0 \\
  l\geq 2:&&[\sigma_{(T)}]=[\sigma_{(LT)}]=[\sigma_{(LL)}]=0\\
  \end{array}\\
  \label{eq:kappes}
  &&\begin{array}{lll}
  l\geq 0:&&[\kappa_{00}]=[\kappa_{(Y)}]=0\\
  l\geq 1:&&[\kappa_{(L)0}]=[\kappa_{(T)0}]=0 \\
  l\geq 2:&&[\kappa_{(T)}]=[\kappa_{(LT)}]=[\kappa_{(LL)}]=0\\
  \end{array}
\end{eqnarray}
are equivalent to (\ref{eq:fpertmc}) and therefore
correspond exactly to the linearized matching conditions in this setting.
Notice that each of the equalities in (\ref{eq:sigmes}) and (\ref{eq:kappes})
is in fact one equation for each $l$ and
$m$ in the appropriate range. We will however refer to them simply 
as equations.

The full linearized matching conditions obviously 
\emph{imply} the following equalities in terms of the
doubly-gauge invariant quantities (\ref{eq:doublies}),
\begin{eqnarray}
  \label{eq:efes}
  &&\begin{array}{lll}
  l\geq 2:&&[f_{00}]=[f]=[f_0]=[f_{(T)}]=0
  \end{array}\\
  \label{eq:kas}
  &&\begin{array}{lll}
  l\geq 1: && [k_{(T)0}]=0\\
  l\geq 2: && [k_{00}]=[k_{(Y)}]=[k_{(L)0}]=[k_{(LL)}]=[k_{(LT)}]=[k_{(T)}]=0.
  \end{array}
\end{eqnarray}
Whether these equations can be regarded as the full set
of linearized matching conditions or not
requires studying their sufficiency, i.e. whether they imply 
(\ref{eq:sigmes})-({\ref{eq:kappes}) or not. This point was not mentioned in
\cite{Mukohyama} and in fact the answer turns out to be negative, although 
in a mild way, as we discuss in the next subsection.


\subsection{On the sufficiency of the continuity of the
doubly-gauge invariants}
\label{sec:suff}

Let us recall that fulfilling the matching conditions requires
finding two $\vec Z^\pm$ such that (\ref{eq:sigmes})-(\ref{eq:kappes})
are satisfied. The key issue for the matching is therefore to show existence
of deformation vectors $\vec Z^\pm$ so that all the equations hold.

A plausibility argument in favour of the sufficiency of
(\ref{eq:efes})-(\ref{eq:kas}) comes from simple equation counting. Indeed,
as already discussed, the linearized matching conditions are spacetime and
hypersurface gauge invariant and therefore can only involve the difference
vector $[\vec T]$, i.e. three quantities. Since constructing
double gauge invariant quantities on each side eliminates this vector,
the number of equations should be reduced exactly by three
if they are to remain equivalent to the original set. This is precisely what
happens as we go from the original forteen equations in
(\ref{eq:sigmes})-(\ref{eq:kappes})
down to eleven equations in (\ref{eq:efes})-(\ref{eq:kas}).
This argument however is not 
conclusive, both because it is not rigorous and
because each equation in those expressions is, in fact, a set of equations
depending on $l$ and $m$, and the range of $l$'s changes with the equations.
Let us therefore analyse this issue in detail. In particular
we need to discuss what are the consequences of the non-existence
of doubly gauge-invariant variables
for $l=0$ and $l=1$ (except for $k_{(T)0}$ which exists for $l=1$), something not
mentioned  in \cite{Mukohyama}.

Let us start by finding explicit expressions for $\sigma$'s valid
in the whole range of $l$'s. As in \cite{Mukohyama}, we decompose
$\gfpert$ in harmonics as
\begin{eqnarray}
  \gfpert_{\alpha\beta} dx^\alpha dx^\beta&=&
  \sum_{l=0}^\infty (h_{ab} Y dx^a dx^b+h_{(Y)}T_{(Y)AB}d\theta^A d\theta^B)\nonumber\\
  &&
  +\sum_{l=1}^\infty
  2(h_{(T)a} V_{(T)A}+h_{(L)a} V_{(L)A}) dx^a d\theta^A \nonumber\\
  &&
  + \sum_{l=2}^\infty
    (h_{(T)}T_{(T)AB}+
    h_{(LT)}T_{(LT)AB}+h_{(LL)}T_{(LL)AB}) d\theta^A d\theta^B,
\label{eq:g1_decomp}
\end{eqnarray}
and $\vec Z$ as
\begin{eqnarray}
  Z_\alpha dx^\alpha
  &=&\sum_{l=0}^\infty z_a Y dx^a+
  \sum_{l=1}^\infty(z_{(T)} V_{(T)A}+z_{(L)} V_{(L)A}) d\theta^A\nonumber\\
  &=&\sum_{l=0}^\infty
  \left(Q Y\fnback -\epsilon N^{-2} z_\lambda Y \bm e_\lambda\right)
  +\sum_{l=1}^\infty (z_{(T)} V_{(T)A}+z_{(L)} V_{(L)A}) d\theta^A,
  \label{eq:Z_decomp}
\end{eqnarray}
which implies
$T_\alpha dx^\alpha = \sum_{l=0}^\infty
(-\epsilon N^{-2} z_\lambda Y \bm e_\lambda)
+ \sum_{l=1}^\infty(z_{(T)} V_{(T)A}+z_{(L)} V_{(L)A}) d\theta^A$.
Inserting these
expressions into (\ref{eq:fpertmc}) and expanding in spherical harmonics
it is straightforward to find
\begin{eqnarray}
  l\geq 0:~~~~~~[\sigma_{00}]=0 &\Leftrightarrow&
  [h_{\lambda\lambda}]+2[Q]N^2\K +
  2 N\partial_\lambda\left(N^{-1}[z_\lambda]\right)=0,\nonumber\\
  l\geq 1:~~~~[\sigma_{(L)0}]=0 & \Leftrightarrow&
  [z_\lambda]+[h_{(L)\lambda}]+
  r^2|_{\supo} \partial_\lambda(r^{-2}|_{\supo}[z_{(L)}])=0,\label{eq:zl}\\
  l\geq 2:~~~~[\sigma_{(LL)}]=0 &\Leftrightarrow &
  [z_{(L)}]+[h_{(LL)}]=0,\label{eq:zL}\\
  l\geq 0:~~~~~[\sigma_{(Y)}]=0 &\Leftrightarrow &
  [h_{(Y)}]+ 2[Q]r^2|_{\supo} \bar\K-
  \epsilon N^{-2}[z_\lambda]\partial_\lambda (r^2|_{\supo}) -\frac{2}{n-2}k_l^2 [z_{(L)}]=0,\nonumber\\
  l\geq 1:~~~~[\sigma_{(T)0}]=0 &\Leftrightarrow &
  [h_{(T)\lambda}]+
    r^2|_{\supo}\partial_\lambda(r^{-2}|_{\supo}[z_{(T)}])=0,\nonumber\\
  l\geq 2:~~~~[\sigma_{(LT)}]=0 &\Leftrightarrow &
  [z_{(T)}]+[h_{(LT)}]=0,\label{eq:zT}\\
  l\geq 2:~~~~~[\sigma_{(T)}]=0 &\Leftrightarrow &[h_{(T)}]=0,\nonumber
\end{eqnarray}
where $[h_{\lambda\lambda}], [h_{(L)\lambda}]$, etc. denote
$e_\lambda{}^a e_\lambda{}^b [h_{ab}], e_\lambda{}^a [h_{(L)a}]$, etc.
Later on we will also write down the explicit expressions for 
(\ref{eq:kappes}) but they are not needed in this subsection.

It is obvious by the form of $f$'s and $\kappa$'s (\ref{eq:doublies})
that the set of equations (\ref{eq:sigmes})-(\ref{eq:kappes})
are equivalent to (\ref{eq:efes})-(\ref{eq:kas}) together with
\begin{eqnarray}
  &&\begin{array}{lll}
  l \geq 2:&&[\sigma_{(L)0}]=[\sigma_{(LT)}]= [\sigma_{(LL)}]=0
  \end{array}
  \label{eq:sigmesToZ} \\
  &&\begin{array}{lll}
  l= 0,1:&&[\sigma_{00}]=[\sigma_{(Y)}]=0 \\
  l= 1:&&[\sigma_{(L)0}]=[\sigma_{(T)0}]=0  \\
  \end{array}
  \label{eq:missing_fs}\\
  &&\begin{array}{lll}
    l= 0,1:&&[\kappa_{00}]=[\kappa_{(Y)}]=0 \\
    l= 1:&&[\kappa_{(L)0}]=0. 
  \end{array}
  \label{eq:missing_ks} 
\end{eqnarray}
Sufficiency of Mukohyama's doubly gauge invariant matching conditions
would follow if these equations  serve
\emph{exclusively} to determine the discontinuity 
$[\vec{T}]$, i.e. 
$[z_\lambda]$ for $l \geq 0$ and $[z_{(T)}], [z_{(L)}]$ for
$l \geq 1$.  Now, the explicit expressions 
(\ref{eq:zl}), (\ref{eq:zT}), (\ref{eq:zL}) show that 
(\ref{eq:sigmesToZ}) determine uniquely  
$[z_\lambda]$, $[z_{(T)}]$ and $[z_{(L)}]$ for $l \geq 2$. So, restricted
to the sector $l \geq 2$ Mukoyama's doubly gauge invariant matching conditions
can be regarded as equivalent to the full set of matching conditions.
Taking all $l$'s into account, however,
the equations turn out \emph{not} to be sufficient. 
To show this,  
it is enough to display one equation involving the discontinuity
of the background metric perturbations and $[Q]$ (but not $[\vec{T}]$) which 
holds as a consequence of the full set of matching conditions
(\ref{eq:sigmes})-(\ref{eq:kappes}) \emph{but not} as a consequence of
(\ref{eq:efes})-(\ref{eq:kas}). Using the fact that
each $l=1$ expression refers to $n-1$ objects (one for each $m$), 
the number of
equations in (\ref{eq:missing_fs})-(\ref{eq:missing_ks}) is
$7n -3$, while the number of unkowns in $[\vec{T}]$ not yet
determined by (\ref{eq:sigmesToZ}), i.e. 
$[z_\lambda]$ for $l=0,1$ and $[z_{(T)}], [z_{(L)}]$ for $l=1$ is $3n-2$, which
is smaller.
It is to be expected, therefore, that
(\ref{eq:missing_fs}), (\ref{eq:missing_ks}) imply
conditions where these variables do not appear.
This can be made explicit, for instance,
by combining $[\sigma_{00}]_{l=0}=0$
with $[\sigma_{(Y)}]_{l=0}=0$ which yields
\[l=0: ~~~~
[h_{\lambda\lambda}]+2[Q]N^2\K +
  2 N\partial_\lambda\left\{\frac{\epsilon N}{\partial_\lambda (r^2|_{\supo})}
  \left([h_{(Y)}]+ 2[Q]r^2|_{\supo} \bar\K\right)\right\}=0,
\]
whenever $\partial_\lambda (r^2|_{\supo})\neq 0$.
(If $\partial_\lambda (r^2|_{\supo})= 0$ it is enough to consider
$[\sigma_{(Y)}]_{l=0}=0$.)
This relation is enough to show that
\emph{the continuity of the doubly-gauge invariant variables of
Mukohyama is not sufficient to ensure the existence
of the perturbed matching.} Of course, this does not invalidate
Mukohyama's approach in any way, which remains interesting and useful. 
It only means that, when using this approach to solve linearized matchings,
one still needs to look more carefully into the $l=0$ and $l=1$ sector to
make sure that the remaining equations (\ref{eq:missing_fs}) and
(\ref{eq:missing_ks}) hold.

On the other hand, equations (\ref{eq:missing_fs}), (\ref{eq:missing_ks})
do not completely determine $[\vec T]$.
The variable $[z_{(T)}]_{l=1}$ only appears in $[\sigma_{(T)0}]_{l=1}=0$,
in the term $\partial_\lambda(r^{-2}|_{\supo}[z_{(T)}])$. As a result,
the matching conditions do not fix $[z_{(T)}]_{l=1}$ completely,
but up to a constant factor times $r^2 |_{\supo}$
(for each $m$).
Recalling that $V_{(T)A}d\vartheta^A$ for
$l=1$ correspond to the three Killing vectors
on the sphere, this arbitrary constant (for each $m$)
accounts for the addition to $[\vec T]$
of an arbitrary Killing vector of the sphere.
This is in accordance with the discussion in
Section \ref{sec:symm_freedom}.
We devote the following subsection to complete the study of
the freedom left in the matching.

\subsection{Freedom in the matching}
\label{sec:freedom_spher}

As already emphasized, solving the linearized matching amounts to 
finding  perturbation vectors $\vec{Z}^{+}$ and $\vec{Z}^{-}$.
Assume now that a linearized matching
between two given backgrounds and perturbations has been done.
It is natural to ask what is the most general matching between those
two spaces, i.e. what is the most general solution for $\vec{Z}^{+}$
and $\vec{Z}^{-}$ of the matching conditions. Geometrically, this
means finding all the possible deformations
of the matching hypersurface $\supo$ which allow the two spaces to be matched.

Since this problem is of interest not only when the full matching 
conditions are imposed but also in situations where layers of matter
are present (e.g. in brane-world or shell cosmologies) so that 
jumps in the second fundamental forms are allowed, we
will analyse this issue in two steps. First, 
we will study the equations involving 
the perturbed first fundamental forms and will determine the
freedom they admit. On a second step we will write down the extra conditions 
coming from the equality of the second fundamental forms.

Thus, let us consider two perturbation configurations
of the same background matching 
and denote their respective sets of difference variables on $\supo$
as $[f]$ and $[f]'$ for any given variable $f$.
Now, we will define the difference between the two configurations
as $<f>\equiv [f]'-[f]$ for any variable $f$.
The assumption that the perturbation on each side is fixed once and for
all implies $<\gfpert>=0$. 
We are assuming that
the linearized matching conditions are satisfied in each case, and so we can
subtract them. Linearity implies that the 
differences of the linearised matching equations become equations
for the difference vector $<\vec Z>$. 
The general solution of these equations clearly determines the freedom
in the deformation of the hypesurface.

The difference of the equations in (\ref{eq:sigmes})
for the two configurations using $<\gfpert>=0$
give the following set of equations
\begin{eqnarray}
  l\geq 0:~~~<\sigma_{00}>=0 &\rightarrow&
  <Q>N^2\K +
  N\partial_\lambda\left(N^{-1}<z_\lambda>\right)=0,
  \label{eq:diff_sigma00}\\
  l\geq 1:~<\sigma_{(L)0}>=0 & \rightarrow&
  <z_\lambda>+
  r^2|_{\supo} \partial_\lambda(r^{-2}|_{\supo}<z_{(L)}>)=0,
  \label{eq:diff_sigmaL0}\\
  l\geq 2:~<\sigma_{(LL)}>=0 &\rightarrow &
  <z_{(L)}>=0,\label{eq:diff_zL}\\
  l\geq 0:~~<\sigma_{(Y)}>=0 &\rightarrow &
  2<Q>r^2|_{\supo} \bar\K-
  \epsilon N^{-2}<z_\lambda>\partial_\lambda (r^2|_{\supo})\nonumber \\
  &&-\frac{2}{n-2}k_l^2 <z_{(L)}>=0,
  \label{eq:diff_sigmaY}\\
  l\geq 1:~<\sigma_{(T)0}>=0 &\rightarrow &
  \partial_\lambda(r^{-2}|_{\supo}<z_{(T)}>)=0,
  \label{eq:diff_sigmaT0}\\
  l\geq 2:~<\sigma_{(LT)}>=0 &\rightarrow &
  <z_{(T)}>=0,\label{eq:diff_zT}\\
  l\geq 2:~~<\sigma_{(T)}>=0 &\rightarrow & 0=0.\nonumber  
\end{eqnarray}
Expressions (\ref{eq:diff_zL}) and (\ref{eq:diff_zT}) readily determine
$<z_{(L)}>_{l\geq 2}=<z_{(T)}>_{l\geq 2}=0$, which substituted in
(\ref{eq:diff_sigmaL0}) give $<z_\lambda>_{l\geq 2}=0$.
As a result, (\ref{eq:diff_sigmaY}) for $l\geq 2$
lead to $<Q>_{l\geq 2}=0$.
Clearly all the equations for $l\geq 2$ are now satisfied.
We now concentrate on the $l=1$ equations.
Equation (\ref{eq:diff_sigmaT0}) implies that
$<z_{(T)}>_{l=1}=a r^2|_{\supo}$, where $a$ is a constant
for each $m$. Combining equations
(\ref{eq:diff_sigma00}), (\ref{eq:diff_sigmaL0}) and (\ref{eq:diff_sigmaY})
for $l=1$ we obtain the following equation for
$r^{-2}|_{\supo}<z_{(L)}>_{l=1}$,
\begin{eqnarray}
  \label{eq:eqforC}
  && \bar\K \partial_\lambda^2(r^{-2}|_{\supo}<z_{(L)}>_{l=1})
  +\left\{(2\bar\K +\epsilon\K)\partial_\lambda (\ln r |_{\supo})
  -\bar\K \partial_\lambda \ln N\right\}
  \partial_\lambda(r^{-2}|_{\supo}<z_{(L)}>_{l=1})\nonumber\\
  &&-\frac{N\K}{r^2}r^{-2}|_{\supo}<z_{(L)}>_{l=1}=0,
\end{eqnarray}
while (\ref{eq:diff_sigmaL0}) and (\ref{eq:diff_sigma00})
determine $<z_\lambda>_{l=1}$ and $<Q>_{l=1}$ respectively (provided 
$\K \neq 0$, which occurs generically).
The two equations for $l=0$ can be rearranged onto
\begin{equation}
  \label{eq:zlam0}
  \bar\K\partial_\lambda(N^{-1}<z_\lambda>_{l=0})
  + N^{-1}<z_\lambda>_{l=0}\epsilon \K\partial_\lambda\ln(r|_{\supo}) =0
\end{equation}
plus the equation (\ref{eq:diff_sigma00}) for $l=0$, which determines
$<Q>_{l=0}$.

Summarizing, we have found that
the freedom in the deformation of the hypersurface compatible with
the linearized matching conditions involving the first fundamental form
is
\begin{eqnarray*}
[\vec Z]'-[\vec Z]=\sum_{l=0}^1
  \left(<Q> Y\vnback -\epsilon N^{-2} <z_\lambda> Y \vec e_\lambda\right)\\
  +a_m \vec V^m_{(T)}+r^{-2}|_{\supo}<z_{(L)}>_{l=1,m} \vec V^m_{(L)},
\end{eqnarray*}
where $r^{-2}|_{\supo}<z_{(L)}>_{l=1,m}$, satisfy (\ref{eq:eqforC}),
$<z_\lambda>_{l=0}$ satisfy (\ref{eq:zlam0})
and the rest of the variables are completely determined as described above.
The term in $a_m$ corresponds to adding Killing vectors on the sphere,
something already discussed in Section \ref{sec:symm_freedom}. 
The rest of terms involve combinations (with functions)
of the conformal Killing
vectors on the sphere and tangential vectors along $\lambda$. 
Notice that the coefficients of the conformal Killing 
(i.e. $<z_{(L)}>_{l=1,m}$
) determine all the rest of the $l=1$ coefficients. In
particular when $<z_{(L)}>_{l=1,m}$ vanishes, then all the $l=1$
terms vanish and the freedom becomes radially symmetric. 

We now add to the analysis
the difference of the equations in (\ref{eq:kappes}).
Due to the fact that all coefficients in $< \vec{Z} >$ vanish for 
$l \geq 2$ we only need to consider the equations for $l=0,1$,
i.e. (\ref{eq:missing_ks}).
We refer the reader to Appendix \ref{sec:app}
for the explicit expressions of (\ref{eq:missing_ks})
in terms of the metric perturbations and $\vec Z$.
For the sake of completeness we also include all the
explicit expressions of (\ref{eq:kappes}) in  Appendix \ref{sec:app}.
The difference of equations (\ref{eq:missing_ks}), see
(\ref{eq:kappa00})-(\ref{eq:kappaY}),
whenever $<\gfpert>=0$ read
\begin{eqnarray}
  &&l=0,1:~~~~~~<\kappa_{00}>=0 \Leftrightarrow\label{eq:diff_kappes1}\\
  &&
  -<Q R^{(\gamma)}_{dbac}>\nback{}^d \nback{}^a e_\lambda{}^b e_\lambda{}^c-\epsilon \partial^2_{\lambda}<Q>
  +\frac{\epsilon}{2 N^2}\partial_\lambda N^2 \partial_\lambda <Q>
  -\epsilon <Q> \K^2 N^2\nonumber\\
  &&
  -\epsilon \K N^2 \partial_\lambda(N^{-2} <z_\lambda>)
  -\epsilon \partial_\lambda(\K <z_\lambda>)=0,\nonumber\\
  &&l=1:~~~~<\kappa_{(L)0}>=0  \Leftrightarrow\\
  &&  -\epsilon\partial_\lambda<Q>+\epsilon\K<z_\lambda>
  +\epsilon<Q>\partial_\lambda\ln (r|_{\supo})
  +r^2|_{\supo}\bar\K\partial_\lambda(r^{-2}|_{\supo} <z_{(L)}>)=0\nonumber\\
  &&l=0,1:~~~~~<\kappa_{(Y)}>=0 \Leftrightarrow \label{eq:diff_kappes3}\\
  && +\frac{1}{2}N^{-2}\partial_\lambda (r^2|_{\supo})
  \left(
    \partial_{\lambda}<Q>
    + \K<z_{\lambda}>
  \right)
  +\frac{1}{2}<Q \nback{}^a \nback{}^b\nabla_a\nabla_b r^2>\nonumber\\
  &&
  -\frac{\epsilon}{2}N^{-2}e_\lambda{}^a<z_\lambda \nback{}^b\nabla_b\nabla_a r^2> +\frac{l(l+n-3)}{n-2}\left\{\epsilon<Q>-2\bar\K<z_{(L)}>\right\}=0.
  \nonumber
\end{eqnarray}
It can be checked that in general 
these equations overdetermine the previous equations, i.e.
(\ref{eq:eqforC}) and (\ref{eq:zlam0}), although
there may be particular cases for which they are compatible.
Therefore, generically, they will imply that
$<z_{(L)}>_{l=1,m}=0$ and $<z_\lambda>_{l=0}=0$, and thence all
the rest of the variables vanish,
$<z_{\lambda}>_{l=1,m}=<Q>_{l=1,m}=0$, $<z_{\lambda}>_{l=0}=<Q>_{l=0}=0$,
so that the only freedom left is given by
\[
[\vec Z]'-[\vec Z]=
  a_m \vec V^m_{(T)}.
\]
Finding in which particular cases equations 
(\ref{eq:eqforC})-(\ref{eq:diff_kappes3}) are
compatible is straightforward but
tedious and will not be carried out explicitly here.


\section*{Acknowledgements}

FM and MM thank CRUP(Portugal)/MCT(Spain) for grant E-113/04. 
FM thanks FCT (Portugal) for grant SFRH/BPD/12137/2003 and CMAT, 
University of Minho, for support.
MM was supported by the projects 
FIS2006-05319 
of the Spanish Ministerio de Educaci\'on y Tecnolog\'{\i}a and 
SA010CO of the Junta de Castilla y Le\'on. 
RV was supported by the Irish IRCSET, Ref. PD/2002/108,
and now is funded by the Basque Government Ref. BFI05.335.

\appendix
\section{Appendix}
\label{sec:app}
For the sake of completeness we devote this appendix to
present the explicit expressions of (\ref{eq:kappes})
in terms of the metric perturbations and $\vec Z$,
which read
\begin{eqnarray}
  &&l\geq 0:~~~~~~[\kappa_{00}]=0 \Leftrightarrow\label{eq:kappa00}\\
  &&\frac{\epsilon}{2}N^2\K[h_{nn}]
  -\frac{1}{2}\nback{}^a e_\lambda{}^b e_\lambda{}^c
  (2\nabla_c[h_{ab}]-\nabla_{a}[h_{bc}])
  -[Q R^{(\gamma)}_{dbac}]\nback{}^d \nback{}^a e_\lambda{}^b e_\lambda{}^c\nonumber\\
  &&-\epsilon \partial^2_{\lambda}[Q]
  +\frac{\epsilon}{2 N^2}\partial_\lambda N^2 \partial_\lambda [Q]
  -\epsilon [Q] \K^2 N^2
  -\epsilon \K N^2 \partial_\lambda(N^{-2} [z_\lambda])
  -\epsilon \partial_\lambda(\K [z_\lambda])=0,\nonumber\\
  &&l\geq 1:~~~~[\kappa_{(L)0}]=0  \Leftrightarrow\label{eq:kappaL0}\\
  &&-\frac{1}{2}[h_{n\lambda}]
  -\frac{1}{2}\nback{}^a e_\lambda{}^b(\partial_b[h_{(L)a}]-\partial_a[h_{(L)b}])
  -\epsilon\partial_\lambda[Q]-\epsilon\K[z_\lambda]\nonumber
  \\
  &&+(\epsilon[Q]+[h_{(L)n}])\partial_\lambda\ln (r|_{\supo})
  +r^2|_{\supo}\bar\K\partial_\lambda(r^{-2}|_{\supo} [z_{(L)}])=0\nonumber\\
  &&l\geq 0:~~~~~[\kappa_{(Y)}]=0 \Leftrightarrow \label{eq:kappaY}\\
  &&-\frac{\epsilon}{2}r^2|_{\supo} \bar\K [h_{nn}]
  +\frac{1}{2}N^{-2}\partial_\lambda (r^2|_{\supo})
  \left(
    \epsilon[h_{n\lambda}]
    + \partial_{\lambda}[Q]
    + \K[z_{\lambda}]
  \right)
  +\frac{1}{2}[\nback{}^a\partial_a h_{(Y)}]\nonumber\\
  &&+\frac{1}{2}[Q \nback{}^a \nback{}^b\nabla_a\nabla_b r^2]
  -\frac{\epsilon}{2}N^{-2}e_\lambda{}^a[z_\lambda \nback{}^b\nabla_b\nabla_a r^2]\nonumber\\
  &&
  +\frac{l(l+n-3)}{n-2}\left\{[h_{(L)n}]+\epsilon[Q]-2\bar\K[z_{(L)}]\right\}=0,
  \nonumber\\
  &&l\geq 1:~~~~~[\kappa_{(T)0}]=0 \Leftrightarrow \nonumber\\
  &&-\frac{1}{2}\nback{}^a e_\lambda{}^b
  (\partial_b[h_{(T)a}]-\partial_a[h_{(T)b}])
  +[h_{(T)n}]\partial_\lambda\ln (r|_{\supo})
  +r^2|_{\supo}\bar\K \partial_\lambda(r^{-2}|_{\supo}[z_{(T)}])=0,\nonumber\\
  &&l\geq 2:~~~~~[\kappa_{(LT}]=0 \Leftrightarrow
  -\frac{1}{2}[h_{(T)n}]+\frac{1}{2}\nback{}^a\partial_a[h_{(LT)}]
  +\bar\K [z_{(T)}]=0,\nonumber\\
  &&l\geq 2:~~~~~[\kappa_{(LT}]=0 \Leftrightarrow
  -\frac{1}{2}[h_{(L)n}]+\frac{1}{2}\nback{}^a\partial_a[h_{(LL)}]
  +\bar\K [z_{(L)}]-\frac{\epsilon}{2}[Q]=0,\nonumber\\
  &&l\geq 2:~~~~~[\kappa_{(T}]=0 \Leftrightarrow
  \frac{1}{2}\nback{}^a\partial_a[h_{(T)}]=0\nonumber.
\end{eqnarray}

\end{document}